\documentclass{tac}
\usepackage{amssymb}
\usepackage{amsfonts}
\usepackage[matrix]{xy}
\usepackage{lscape}
\input diagxy
\usepackage{comment}
\newtheorem{defi}{Definition}
\newtheorem{teo}{Theorem}

\newcommand{\f}[2]{\frac{\displaystyle #1}{\displaystyle #2}}
\def\sq{\sqrt}
\def\sq2{\sqrt{2}}
\def\sq12{\sq{12}}
\def\dsq2{\f{1}{\sqrt{2}}}
\def\be{\begin{equation}}
\def\ee{\end{equation}}
\def\q={\quad = \quad}
\def\lra{\longrightarrow}
\def\Lra{\Longrightarrow}

\def\A{{\mathbb A}}
\def\B{{\mathbb B}}
\def\C{{\mathbb C}}
\def\D{{\mathbb D}}
\def\E{{\mathbb E}}
\def\Cat{{\mathbb Cat}}

\def\Sammy{Sammy}

\def\N{{\mathbb N}}

\def\W{{\mathbb W}}
\def\X{{\mathbb X}}
\def\Y{{\mathbb Y}}

\def\Z{{\mathbb Z}}

\def \so{{\bf 1}}
\def\st{{\bf 2}}
\def\c{{\bf c}}
\def\f{{\bf f}}
\def\n{{\bf n}}
\def\sto {\overline{\st}} 
\def\omeg{{\mathbb \omega}}

\newcounter{examnum}[section]
\newcounter{remarnum}[section]
\begin{document}
\title{Computability and Complexity of Categorical Structures}
\author{Noson S. Yanofsky}\address{Department of Computer and Information Science,
Brooklyn College, The City University of New York, Brooklyn, N.Y. 11210.} 
\eaddress{noson@sci.brooklyn.cuny.edu.} \thanks{ Support for this project was provided by a PSC-CUNY Award, jointly funded by The Professional Staff Congress and The City University of New York. This work was also supported by a generous``Physics of Information'' grant from The Foundational Questions Institute (FQXi). }
\maketitle
\begin{abstract}
\noindent We examine various categorical structures that can and cannot be constructed. We show that total computable functions can be mimicked by constructible functors. More generally, whatever can be done by a Turing machine can be constructed by categories. Since there are infinitary constructions in category theory, it is shown that category theory is strictly more powerful than Turing machines. In particular, categories can solve the Halting Problem for Turing machines. We also show that categories can solve any problem in the arithmetic hierarchy. 
\end{abstract}

\section{Introduction}
Categories are used in almost every branch of mathematics and theoretical computer science. As of late, a large part of theoretical physics also use categorical structures.  There are, however, parts of mathematics that have not been conducive to categorical thinking. 
We are interested in what exactly can and cannot be constructed or computed using category theory? What is within its ability to describe and what is outside its ability? We shall use the words ``constructed,'' and ``computed'' interchangeably throughout this paper.

In the 1930's mathematicians asked what processes are computable and what processes are not. Although there was much intuition, there was no clear way to answer this question until there was a concrete model of computation. Alan Turing introduced the idea of a Turing machine in order to formulate the intuitive notion of a computation. By fixing our ideas with one particular  model, he was able to move the discussion forward. In a similar vain, if we are going to move forward investigating what can and cannot be described by categories, we are going to need a concrete  model of categorical constructions. 

In \cite{KCC} I formulated a programming language that described categories, functors and natural transformations. The programing language was named {\it Sammy} in honor of Sammy Eilenberg who was one of the founders of category theory and who also had a deep interest in formal language theory and computability theory. Sammy has certain constants and its operations are simple categorical constructs (like composition of functors and Kan extensions of functors) that are used in everyday mathematical reasoning.We will examine what categorical structures can be constructed by a Sammy program and what structures cannot. 

In classical computability theory we are not only interested in what is computable but also in what is {\it relative computable}, that is, what is computable given other information. In this paper, we are not only interested in what structures are constructible but also in what structures are  {\it relative constructible}. We will examine what structures are constructible when other structures are used in a Sammy program as inputs. As a simple example, it might be hard to construct the category of Lie groups out of the constants in the Sammy programing language, but it is not hard to construct if you are {\it given} the category of smooth manifolds. Lie groups are simply group objects in the category of smooth manifolds.      

Of course this only tells us what can and cannot be described by Sammy. What about other ways of describing categorical structures? There is a similar problem in classical computability theory where the Turing machine is the computational model. This problem is dealt with by the Church-Turing thesis which states that everything that can be computed by any other computational model can also be computed by a Turing machine. More important is the contrapositive of this thesis which says that whatever cannot be computed by a Turing machine cannot be computed by any other computational model. We believe an unprovable thesis that says that whatever can be constructed categorically from basic terms can be constructed by a Sammy program.

Sammy rests heavily on the important work  \cite{CCFOM} of William Lawvere.  In order to  use category theory as a foundation of mathematics, he formulated first order axioms for the theory of categories. We are using categories as a way of describing structures and we use a programming language that has some similarity to those axioms. Just as the axioms consist of small categories that generate all of $\Cat$, so too our programming language describes operations and small categories that generate constructable categories.

Some of our proofs of the fact that certain structures will not be constructible will be very similar to proofs of certain limitations in classical computability theory and classical Kolmogorov complexity theory. Usual  computability limitations are based on variations of the self-referential  liar paradox (``This statement is false''). G\"odel's famous incompleteness theorem (``This statement is unprovable'') and Turing's unsolvability of the halting problem (``This program will output the wrong answer when asked if it will halt or go into an infinite loop'') are two such examples  (see \cite{Universal} and \cite{OLR} for comprehensive surveys of such limitations.) In contrast, the limitations of Kolmogorov complexity 
are based on the Berry Paradox: consider the number described by ``The least number that needs more than fifteen
words to describe it.'' This sentence has twelve words. That is, there is a description of a number that is shorter than it is supposed to be. We will have a categorical version of such a limitation. 

This work grew out of our work in the Kolmogorov complexity of categories \cite{KCC, AITinCA, ECC}. Classical Kolmogorov complexity describes how complicated a string is by looking at the shortest program that describes that string. If a short program exists to describe a string, then the string does not contain much information and is compressible. In contrast, if only a large program can describe a string, than it has a lot of information. If the only way to describe a string is to literally have it printed out, then it is extremely complex or, in other words, random. In \cite{KCC} and in forthcoming papers we generalize classical Kolmogov complexity by looking at categorical structures as opposed to strings. In that work we needed the Sammy programming language to describe categorical structures. We were interested in the shortest Sammy program which is a description of that structure. In this work we use that programming language to decide what is and is not constructible.  (Knowledge of \cite{KCC} is not necessary to read this paper.)

Another way to approach this work is via hyper-computation. Turing was one of the first people to formulate a concrete notion of computation with his Turing machine. Over the years others have formulated other notions. Most of these other notions were all shown to be equivalent to Turing machines and hence each other. There are, however, some notions of machines that are more powerful than Turing machines. A prominent examples is Joel Hamkins and Andy Lewis's notion of an infinite time Turing machine. These are machines that can do an infinite amount of operations (see, e.g., \cite{Hamkins}). Another example is the Blum-Shub-Smale machine which can deal with the ring of real numbers as opposed to the usual natural 
numbers \cite{BSS}. These models of computation, and several others, are more powerful than standard Turing machines.
We look at category theory as a more powerful computational tool. Since it is used in standard mathematics and there are infinitary operations, one should expect it to do more. This paper describes some of the more powerful operations that it can perform. 

An analogy is in order. G\"{o}del's incompleteness theorem tells us that for Peano arithmetic, there are some statements that are true but not provable within that system. In what sense are they true? They are true in the model of natural numbers, or another way to say this is that they are true in more powerful ZFC set theory. What about classical computability theory? For what larger, more powerful systems can the halting predicate be constructed? Category theory is such a system. We show that parts of classical computability theory can be worked out within category theory. 
   
\vspace{.2in}

This paper is organized as follows: Section 2 describes the Sammy programming language and has a mini-appendix on Kan liftings, one of the operations in the programing language. Section 3 gives several examples of structures that are constructible, including some that cannot be done by regular computers. The next section discusses several constructions that are beyond the ability of Sammy. Section 5 shows the power of Sammy within the  arithmetic hierarchy. The next section discusses  some possible ways this work can progress in the future. The paper concludes with an appendix containing the formal syntax and semantics of Sammy.

\vspace{.2in}

\noindent{\bf Acknowledgment.} I thank Michael Barr, Gershom  Bazerman, John Connor, Joel Hamkins, Florian Lengyel, Fred Linton, Dustin Mulcahey, Robert Par\'{e}, Andrei Rodin, Robert Seely, Ross Street, David Spivak, Jouko V\"{a}\"{a}n\"{a}nen, Gerald Weiss, Neng-Fa Zhou, and two anonymous reviewers for helpful suggestions and discussions.  

\section{Describing Categorical Structures}
In order to fix our ideas, we formulate a programming language named Sammy that can be used to describe categorical structures. Most of the lines of the programming language will consist of  an assignment statement that sets some variable equal to a categorical construction
$$\W = \verb"Operation"(\X, \Y, \Z).$$
The variables $(\W, \X, \Y, \Z)$ can contain categories, functors, or natural transformations. They are set to constants or constructions  built by operations on previously defined variables. 
The operations are typical operations that are used in the description of categorical structures to be described below. Some lines in the program will also have conditional breaks to control the execution of the program. 

 The constant categories of Sammy are $\bf 0$, the empty category, $\bf 1$,  the category with one object, $0$, and $\bf 2$, the category $0\lra 1$ with two objects and one nontrivial morphism between them. We will also need the constant category \verb"Cat" which corresponds to the category of all small categories.  Constant functors are the unique morphisms $!_{\bf 01}:\bf \bf 0 \lra \bf 1$, $!_{\bf 02}:\bf \bf 0 \lra \bf 2$,
$!_{\bf 0Cat}:\bf 0 \lra \verb"Cat"$, $!_{\bf Cat1}: \verb"Cat" \lra \bf 1$, and $!_{\bf 21}: \bf 2 \lra \bf 1$. There are also identity constant functors on  $\bf 0, \bf 1,\bf 2$ and \verb"Cat". Sammy's only constant natural transformations are identity natural transformations on the constant functors.

There are several operations that take a single input. For a functor $F:\A \lra \B$ if we set 
$\C=\verb"Source"_1(F:\A \lra \B)$ then $\C=\A$. That is, $\verb"Source"_1$ takes a functor and outputs the category that is the source of the functor. There is a similar operation $\C=\verb"Target"_1(F:\A \lra \B)$. There are similar operations for natural transformations $\verb"Source"_2$ and $\verb"Target"_2$. For a given category $\A$, the operation
$F=\verb"Ident"_0(\A)$ has the effect that $F=Id_\A$.There is a similar operations $\verb"Ident"_1$ for functors.  For a category $\A$, if we let $\C=\verb"Op"_0(\A)$ then $\C=\A^{op}$. The $\verb"Op"_1$ operation works on functors.

One of the three main operation in categories is composition. For functor $F:\A \lra \B$ and $G:\B \lra \C$ we let $H=\verb"Comp"(G,F)$ be the composition of the two functors. If the target of the $F$ is not the same as the source of $G$ then the Sammy program will not be defined.  
For natural transformations of the 
appropriate source and target Sammy has a horizontal composition and vertical composition written as  $\alpha=\verb"Hcomp"(\beta, \gamma)$ and  $\alpha=\verb"Vcomp"(\beta, \gamma)$. For categories $\A$ and $\B$, we will let $\C = \verb"Pow"_0(\A, \B)$ be the category of all functors and natural transformations from $\A$ to $\B$. The operation $\verb"Pow"_1$ is  defined on functors and takes $F:\A \lra \B$ and $G:\C\lra \D$ and outputs the induced functor 
$$G^F=G\circ - \circ F: \C^\B \lra \D^\A.$$

Probably the most important operations are the Kan extensions. For categories $\A, \B$ and $\C$ and functors $G:\C \lra \A$ and $F:\C \lra \B$, a right Kan extension of $F$ along $G$ is a pair $(R, \alpha)=\verb"KanEx"(G, F)$
where $R:\A \lra \B$ and $\alpha:R\circ G \lra F$ is a natural transformation that satisfies certain universal properties. A left Kan extension can be described using the $\verb"Op"_0$ operation on $\B$. A Kan extension induces a natural transformation whenever another functor and natural transformation are given: for every $H:\A \lra \B$ and $\beta: H \circ G \lra F$ there is a unique $\gamma=\verb"KanExInd"(G,F, H, \beta)$
where $\gamma:H \lra R$ and satisfies $\alpha \cdot \gamma_G = \beta$.  
The uniqueness of $\gamma$ basically says that the information in $\beta$ is the same as the information in $ \gamma$.

Using Kan extensions one can derive, products, coproducts, pushouts, pullbacks, equalizers, coequalizers, (and constructible) limits, colimits, ends, coends, etc. It is a well-known fact that if $G:\A \lra \B$ is a right adjoint (left adjoint, equivalence, isomorphism), then its left adjoint (right adjoint, quasi-inverse, inverse) $G^*:\B \lra \A$ can be found as a simple Kan extension of the identity $Id_\A$ along $G$, that it, $G^*=\verb"KanEx"(G, Id_{\A})$.

The left and right Kan extension along the unique constant functors
$$
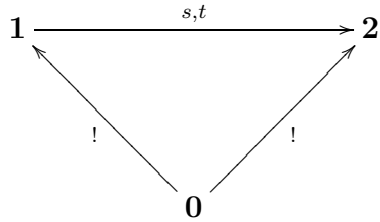
\xymatrix{{\bf 1} \ar[rrrr]^{s,t}&&&& {\bf 2}\\ \\ && {\bf 0}\ar[lluu]^{!} \ar[rruu]_{!}}$$
give the functors $s:\bf 1 \lra \bf 2$ and $t:\bf 1 \lra \bf 2$ that picks out the source and target of $\bf 2$. 

In addition to Kan extensions Sammy also has right and left Kan liftings. Since these are less well known, we will have a mini-appendix about them at the end of this section. Briefly, for categories $\A, \B$ and $\C$ and functors $P:\B \lra \C$ and a functor $F:\A \lra \C$, a right Kan lifting of $F$ along $P$ is constructed as  
$ (R, \alpha)=\verb"KanLif"(P, F) $
where $R: \A \lra \B$ and $\alpha :P \circ R \lra F$ is a natural transformation which satisfies certain universal properties. There is also an induced Kan lifting natural transformation,that is, if we have a $H: \A \lra \B$ and a $\beta: P \circ H \lra F $ then we have an induced  
$ \gamma=\verb"KanLifInd"(P,F, H, \beta) $
where $\gamma: H \lra R$ and satisfies $\gamma = \alpha \circ P (\beta)$.  A left Kan lifting can be described using the $\verb"Op"_0$ operation on $\B$.


One can see these three operations --- composition, extension and lifting --- as three sides of a triangle:
$$
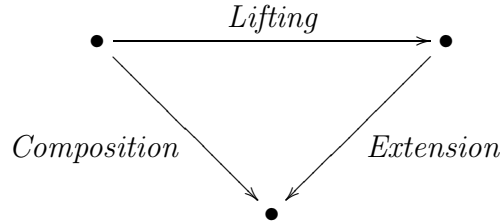
\xymatrix{ {\bullet } \ar[rrrr]^{\textit{Lifting}} \ar[rrdd]_{\textit{Composition}}&& && { \bullet}\ar[ddll]^{\textit{Extension}}
\\ \\
&& { \bullet} 
}$$ 
Each side demands the other two sides as input. That is,  Lifting needs Composition and Extension as input. Extension needs Lifting and Composition as input. And Composition needs Extension and Lifting as input. Another way to see this is to realize that given any two functors that share a domain or a codomain or one domain is the other's codomain, there is an operation that describes a third functor that connects them.  

Another operation that will be needed is the actual composition operation of a category, $ \verb|MapComp| $.
 For every category $\C$, there is a functor $\verb|MapComp|: \C^{\st}\times_{\C^{\so}} \C^{\st} \lra \C^{\st}$ that takes pairs of composible arrows to their composition. 

Programs will not just be sequences of statements. Some  lines will be given labels. There will be conditional statements of the form $$\verb"If " \alpha \verb" ==  "\beta \verb" goto " L$$ where $\alpha$ and $\beta$ are natural transformations and $L$ is a label of some line. When such a conditional statement is met, the program will evaluate to see if the two natural transformations are exactly the same and if so, execution of the program will go to the first line of the program with the label $L$. This will permit us to control the execution of the programs. These conditional statements can be used to not only compare natural transformations but also functors and categories.

The last line of any program will be a $\verb|Return|$ operation, such as $\verb"Return"(\X, \Y, \Z)$ where $\X, \Y$ and $\Z$ are three categorical structures.   This will tell what structure are being defined by the program. There can be any finite number of categorical structures returned. In general, there are many parts to a structure. For example, a monad/triple is a category, an endofunctor and a few natural transformations that satisfy certain requirements. 

These are the major operations needed to describe categorical structures. The syntax and semantics of Sammy is presented in an appendix at the end of this paper.

There are a few remarks that need to be stated about Sammy:
\begin{itemize}
\item Numbers, strings, trees, graphs, arrays, and other typical data types are not mentioned in our programming language.  The only types that Sammy deals with are
categories, functors, and natural transformations. This was done on purpose. We want to emphasize that the usual data types can be derived from the categorical structures. Categories and algorithms are more ``primitive'' than numbers, strings, etc. 

\item This is not the first time a programming language has been formulated to describe categorical structures.  An important work in this area is {\it Computational Category Theory} by Rydeheard and Burstall \cite{RandB}. Another work found in the literature is Tatsuya Hagino's thesis \cite{Hagino}. These languages are, however, different than Sammy. These programming languages were intended to be implemented and to get computers to actually calculate with categories. In contrast, Sammy is used to describe structures. As we shall see, Sammy cannot be implemented.

\item Sammy might not the best or smallest programming language that can describe categories. We could have used fewer operation. For example, $\verb"Target"_1$ could have been derived using $\verb"Source"_1$ and $\verb"Op"_1$. 
Since categories and functors are special types of natural transformations, we could have formulated most of our programming language with the single type of natural transformations. This, however, would have made Sammy programs unreadable. Our goal is neither succinctness nor cleverness. Rather, our goal is clarity. 

\item Sammy can never be totally implemented. While one can write compilers for such a language, the compilers will not always work. This can be seen when we prove later that Sammy can construct a functor that essentially solves the halting problem. Turing taught us that no computer can solve the halting problem. We can conclude from this that while Sammy can be partially implemented, we will never get any computer to run every Sammy program. (This is in line with similar statements about infinite-time Turing machines and other super-computable models of computation.)

\item There is a need for a Church-Turing type thesis. The classic Church-Turing thesis says that whatever can be computed, can be computed by a Turing machine. We need a thesis that says that whatever can be constructed by categorical means, can be constructed using the Sammy programming language. We shall show this in the same way that it is shown with Turing machines, namely, we will show that many structures that one believes should be constructible, can be constructed by Sammy. Alas, this is a thesis and not a theorem because we cannot exactly characterize what can and cannot be constructed by categorical means. 

\item We use the term ``relative computable'' or ``relative constructible'' for programs that use already defined structures. For example, a program that describes the category of Lie groups might use the category of smooth manifolds as a given category. 

\item Just as hardly anyone ever actually sits down and writes up the code for a Turing machine, so too, we have no intention to write much Sammy code to describe various constructions. Such programs would be too long, too complicated and not very enlightening. We do write some pseudo-code for Sammy programs in this paper and we will show how different constructions are made in typical category theoretic  notation. 

\item Kan extensions (and Kan liftings) are only defined up to isomorphism. Hence, products, coproducts, equalizers, colimits, etc are also only defined up to isomorphism. We are left with the question as to which of the many possible Kan extensions does Sammy actually give. We leave this as implementation-dependent. Let the virtual machine choose one of them. Categorically they are all isomorphic and indistinguishable. 

\item In \cite{KCC} we already showed how to construct equalizers, coequalizers, pullbacks, pushouts, comma categories, slice categories, the category $\sto$ which is the category with two distinct objects and an isomorphism between them, and several other constructs. We will not repeat those constructions here.   

\item Not every operation will work. Sometimes a program might want to compose two functors that are not adjacent to each other, or take a Kan extension that does not exist. In that case the Sammy program will just not be defined. This is similar to a case in classical computers where a program might come to divide by zero. Such programs are not valid. As in classical computers, it is  in general impossible to syntactically determine if a Sammy program is valid.

\item If an operation is defined, then it is assumed to take a finite amount of time to execute. This is true despite the fact that the operation might be taking a limit over an infinite category or deciding if two natural transformations with an infinite class of components are equal. Every program either goes into an infinite loop or halts after a finite amount of time. Programs do not go on for a transfinite amount of time. With this fact in mind, we can go on and talk of the computational complexity of a categorical construct. 
 
\item There are about 20 types of commands in the Sammy programming language and the programs are of finite length. This makes it very easy to encode and decode a Sammy program as a finite number. For Sammy programs, a number will correspond to a functor $P: \so \lra \omeg$ where $\omeg$ is the totally ordered category of natural numbers. We will have Sammy programs that describe functors which could be describe the number of a Sammy program. This will give us the self-reference that is so central to computability theory. 
\end{itemize}

Formally, we say that a category $\A$ is {\it constructible} if there exists a Sammy program that $\verb"Return"$s a category that is isomorphic to $\A$. There is a similar definition for functors. A natural transformation $\alpha$ is {\it constructible} if there exists a Sammy program that returns $\alpha$ with the command $\verb"Return"$s $(\alpha)$.

\vspace{.2in}

Although Kolmogorov complexity of categories will not be the central focus of this paper, several such concepts will be used and need to be  spelled out. When a certain structure is constructible by Sammy, we will be interested in what is the smallest number of Sammy operations needed to describe this structure. For a structure ${\X}$, its Kolmogorov complexity, $K_{Sammy}({\X})=K({\X})$ is the smallest number of operations in a Sammy program that can produce ${\X}$. For two categorical structures ${\X}$ and ${\Y}$, the relative Kolmogorov complexity, $K_{Sammy}({\X}|{\Y})=  K({\X}|{\Y})$ is the smallest number of Sammy operations needed to describe ${\X}$ given that ${\Y}$ can be used as  input to the program.

If there is a finite number of operations so that one can go from one categorical structure to another and vice versa, we say that the Kolmogorov complexity of these categorical structures are approximately the same. In detail, if there exists a $c$ such that for all appropriate categorical structures, $\X$ and $\X'$, one can change $\X$ to $\X'$ and vice versa in $c$ Sammy operations, that is $|K(\X) - K(\X')|\leq c$, then we write $K(\X) \approx K(\X')$. As an example, notice that only one Sammy operation is needed to go from category $\A$ to functor $Id_{\A}$ and vice versa. Hence $K(\A) \approx K(Id_{\A})$. 

There is a need for something called an {\it invariance theorem}. The Kolmogorov complexity does not depend on the programing language that is used to describe the structures. Imagine another programming language to describe categorical structures called {\it Saunders} (after the other founder of category theory, Saunders Mac Lane.) Then since presumably both languages can program any constructible categorical structure, they can each program the other's operations. That means there exist compilers that can translate Sammy programs into Saunders programs and vice versa. From this, we can prove the following theorem: 
There exists a constant $c$ such that for all categorical structures $\X$ we have $|K_{Sammy}(\X) - K_{Saunders}(\X)|\leq c$.
The proof of this essentially lets $c$ be the larger of the two possible compilers: The Sammy compiler in the Saunders language or the Saunders compiler in the Sammy language. By precomposing the appropriate compiler to a program in one language we get the program in the other language.    

\vspace{.2in}

\noindent {\bf A mini-appendix on Kan liftings. }
 In this appendix we familiarize ourselves with the basic properties and ideas of Kan liftings.  Kan extensions are well known and as Saunders Mac Lane proclaimed \cite{CWM} ``All concepts are Kan extensions.'' Their cousins, Kan liftings, are less well known but we believe just as ubiquitous. While Kan extensions are found everywhere in the literature, Kan liftings are only found in a few places. 
For the prehistory of Kan liftings see \cite{Ulmer}. Kan liftings and their relationship with relative adjoints are developed further in 
\cite{Street1} and 
\cite{Street2}.

Let $\A, \B$ and $\C$ be three categories and let $P: \B \lra \C$ be any functor. $P$ induces a functor $$P\circ -=P^\A:\B^\A \lra \C^\A.$$ Left and right adjoints of $P^\A$, if they exist,  are called left and right Kan liftings:
$$\xymatrix{\B^\A \ar[rrr]^{P \circ -}&&& \C^\A. \ar@/_.3in/[lll]_{L_P}^\bot \ar@/^.3in/[lll]^{R_P}_\bot} $$
In detail, for $F\in \C^\A$ and $H \in \B^\A$ we have 
$$\B^\A(L_P(F) ,H )\cong\C^\A (F,P \circ H) \qquad \mbox{and} \qquad \B^\A(H ,R_P(F)  )\cong \C^\A (P\circ H,F).$$

One can restate this in terms of universal properties. The following diagrams will be helpful: 
$$ \xymatrix{\A \ar[rrrr]_H \ar[ddrr]_F\ar@/^.2in/[rrrr]^{L_P(F)}_{\Downarrow \beta} &&&& \B\ar[ddll]^P && \A \ar[rrrr]_{R_P(F)} \ar[rrdd]_F\ar@/^.2in/[rrrr]^H_{\Downarrow \beta} &&&&\B \ar[ddll]^P
\\ &&\Longrightarrow \alpha &&&&&&\Longleftarrow \alpha
\\ 
&&\C&&&&&&\C
}$$

A left Kan lifting of $F$ along $P$ is a functor $L_P(F)$ and a natural transformation $\alpha:F \Lra  P \circ L_P(F) $ which is universal from $F$ to $P\circ -$. This means that for any other $H:\A \lra \B$ and $\beta : F \Lra P \circ H$ there is a unique natural transformation $\gamma: L_P(F) \Lra H$ such that 
$$\xymatrix{
F  \ar@{=>}[ddrr]_\beta  \ar@{=>}[rr]^\alpha && P \circ L_P(F)  \ar@{=>}[dd]^{P \circ \gamma}&& L_P(F) \ar@{=>}[dd]^\gamma
\\
\\
     && P \circ H               && H
}
$$
That is,
$\beta =  (P \circ \gamma)\circ \alpha.$

 A right Kan lifting of $F$ along $P$ is a functor $R_P(F)$ and a natural transformation $\alpha: P \circ R_P(F) \Lra F$ which is universal from $P\circ -$ to $F$. This means for any $H:\A \lra \B$  and $\beta:  H \Lra R_P(F)$, there is a unique natural transformation $\gamma :P \circ H \Lra F$ such that

$$\xymatrix{R_P(F) && P \circ R_P(F) \ar@{=>}[rrr]^{\alpha} &&&  F
\\
\\
 H\ar@{=>}[uu]^\beta && P \circ H   \ar@{=>}[uu]^{(P \circ \beta)}\ar@{=>}[rrruu]_{\gamma}          
}$$
That is,
$\gamma = \alpha \circ (P \circ \beta ). $

Kan liftings can be constructed with limits and colimits. First some preliminaries.  
For $P:\B \lra \C$,  $F: \A \lra \C$ and object $a$ in $\A$  consider the  comma category $(F(a)\downarrow P)$ which consists of  arrows
$f:F(a) \lra P(b)$  in $\C$ for some $b$ in $\B$. The morphisms of this category consists of commuting triangles 
$$\xymatrix{&F(a)\ar[ld]_f\ar[rd]^{f'}\\
P(b)\ar[rr]_{P(g)} &&P(b')
}$$
for some $g:b \lra b'$ in $\B$. This category has an obvious forgetful functor 
$Q_a: (F(a)\downarrow P) \lra \B$ which sends $f:F(a) \lra P(b)$ to $b$ and sends the above diagram to the $g:b \lra b'$ arrow  in $\B$. Any $h:a \lra a'$ in $\A$ induces a functor $$h^*:(F(a')\downarrow P)\lra (F(a)\downarrow P)$$ which commutes with the forgetful functor: $Q_{a'} =  Q_{a}h^*$. This means that whatever is in the image of $Q_{a'}$ will also be in the image of $Q_{a}$.

Kan extensions exist and can be described with limits and colimits when the target category has the proper limits and colimits. No requirements are made for the functors between the categories. In sharp contrast to this --- and analogous to the algebraic topological notions ---  Kan liftings only exist when $P$ is a continuous functor (and when $\B$ has appropriate limits). 

 If the  left Kan lifting exists, it can be  calculated on the object $a \in \A$ as 
$$L_P(F)(a)=Lim (Q_a:(F(a)\downarrow P) \lra \B).$$
We have many steps in showing that this formula works. First, for any $h: a \lra a'$ we have that $Q_{a'} =  Q_{a}h^*$ we can see that there is an induced $L_P(F)(a) \lra L_P(F)(a')$ which shows that $L_P(F)$ is functorial. 

The Definition of $\alpha:F \Lra P \circ L_P(F)$ is given as follows. A typical limit looks like the following diagram in $\B$:
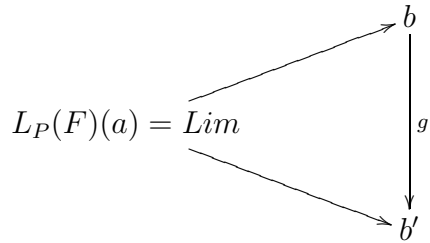
$$\xymatrix{
&&b \ar[dd]^g
\\
L_P(F)(a)=Lim\ar[rru]\ar[rrd]
\\
&& b'
}$$
Apply $P$ to this diagram and notice that by the continuity of $P$, we have that $P(Lim)=Lim(P)$ we get the following diagram in $\C$
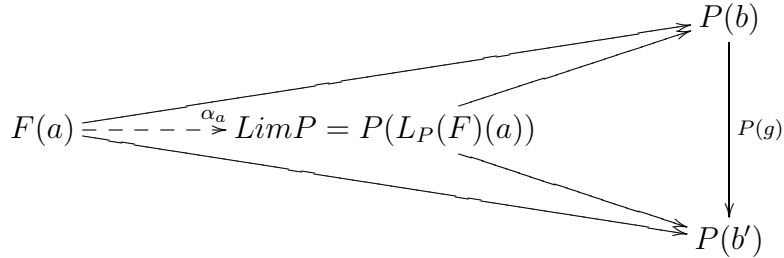
$$\xymatrix{
&&&&P(b) \ar[dd]^{P(g)}
\\
F(a)\ar@{-->}[rr]^{\alpha_a}\ar[rrrru]\ar[rrrrd]&&Lim P=P(L_P(F)(a))\ar[rru]\ar[rrd]
\\
&&&&P(b')
}$$
Since the $F(a)$ also has arrows to everything in the limit, there is an induced $\alpha_a$. Naturality of $\alpha$ can be seen by looking at the following commutative square:
$$
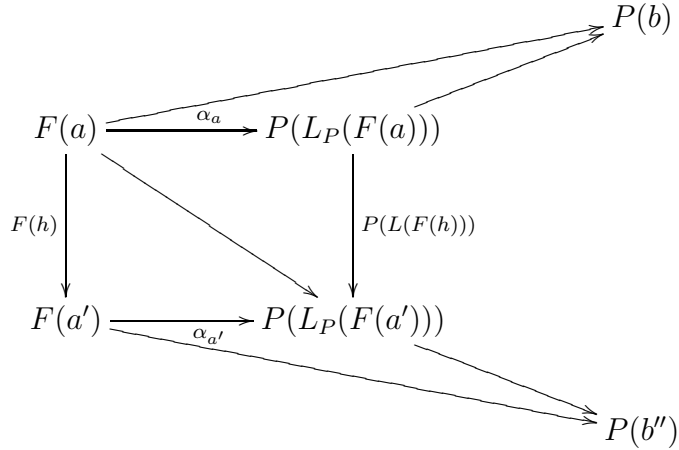
\xymatrix{
&&&&P(b)
\\
F(a)\ar[rr]^{\alpha_a}\ar[dd]_{F(h)}\ar[rrdd] \ar[rrrru]&&P(L_P(F(a)))\ar[dd]^{P(L(F(h)))}\ar[rru]
\\
\\
F(a')\ar[rr]_{\alpha_{a'}}\ar[rrrrd] &&P(L_P(F(a')))\ar[rrd]
\\
&&&&P(b'')
}$$
(The diagonal map is induced by the limit.)

The universal property of $\alpha$ is derived as follows. Let $H: \A \lra \B$ and $\beta:F \Lra P \circ H$. This means that for every $a$ in $\A$ there is a map $\beta_a:F(a) \lra P(H(a))$. This map is necessarily in $(F(a)\downarrow P)$ which will look like
$$\xymatrix{
&&&&P(b) \ar[dd]^{P(g)}
\\
F(a)\ar[rrrru]\ar[rrrrd]_{\beta_a}&&
\\
&&&&P(H(a))
}$$
Taking the forgetful functor, then the limit and then the $P$ functor gives us this:
$$\xymatrix{
&&&&P(b) \ar[dd]^{P(g)}
\\
F(a)\ar[rr]^{\alpha_a}\ar[rrrru]\ar[rrrrd]_{\beta_a}&&Lim P=P(L_P(F)(a))\ar[rru]\ar[rrd]_{P\gamma_a}
\\
&&&&P(H(a))
}$$
The induced $\gamma_a$ will come from the fact that $H(a)$ is in the image of the forgetful functor. Naturality of $\gamma$, which means the commutativity of the following square:
$$\xymatrix{
L_P(F(a))\ar[rr]^{L_P(F(h))}\ar[dd]_{\gamma_a}\ar[rrdd]&& L_P(F(a'))\ar[dd]^{\gamma_{a'}}
\\
\\
H(a) \ar[rr]_{H(h)}&& H(a')
}$$
We leave this for the reader. 

Notice that if for a given $a$ in $\A$ there exists a $b$ in $\B$ such that $F(a) = P(b)$ then $id:F(a) \lra P(b)$ will be an initial object in $(F(a) \downarrow P)$. The limit of the forgetful functor $Q_a$ will then be isomorphic to $b$. Notice that if there are several $b$'s that have this property, then they can all be the limits because they will all be isomorphic. 
 
The right Kan lifting is done similarly with the category $(P \downarrow F(a))$ and a similar forgetful functor $Q'_a:(P \downarrow F(a)) \lra \B$. The formula will be 
$$R_P(F)(m)=Colim(Q'_a:(P \downarrow F(a)) \lra \B).$$
$R_P(F)$ exists when $\B$ is cocomplete, and $P:\B \lra \C$ preserves colimits.

\section{Constructible Categorical Structures}

We begin by forming a few little constructions that will be needed for the rest of the paper.
Since classical computability and complexity theory is based on the natural numbers, we begin with their construction. 
The coequalizer
$$ \xymatrix{ \so \ar@<-.1in>[r]_t \ar@<.1in>[r]^s & \st \ar[r]^{\rho}& \N}$$
gives the (infinite) natural numbers as a monoid. That is, $\N$ is the one-object category whose morphisms are the natural numbers. The single object of $\N$ is $\ast$. The totally ordered category of natural numbers is formulated as a coslice category of the one object category. 
$$\ast / \N  = \omeg =  \qquad 0 \lra 1 \lra 2 \lra 3 \lra \cdots. $$

In order to get the discrete category of natural numbers we need a special type of categorical structure which we will call a {\it iso-comma category}. Whereas for functors $L:\A \lra \C$ and $R:\B \lra \C$ the objects of a comma category are morphisms of the form $f :L(a) \lra R(b)$ in $\C$, here we need isomorphisms.  We can construct the iso-comma category, denoted $L \downarrow \uparrow R$ from the following pullbacks:
$$
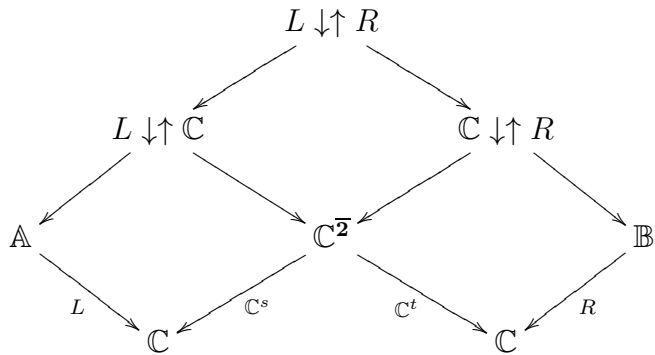
\xymatrix{&& L \downarrow \uparrow R \ar[ld]\ar[rd] \\
 & L \downarrow \uparrow \C \ar[rd]\ar[ld] &   &\C \downarrow \uparrow R \ar[rd]\ar[ld]&\\
\A\ar[rd]_L&&\C^{\sto}\ar[rd]_{\C^t}\ar[ld]^{\C^s}&&\B\ar[ld]^R \\
 &\C && \C
}$$ 
Using this construction we can get {\it iso-slice categories} $\C\downarrow \uparrow c$ and {\it iso-coslice categories} $c\downarrow \uparrow\C$. The obvious forgetful functors from $L \downarrow \uparrow R$ to $\C$ work as usual.

 The discrete category of natural numbers is described as
$$\ast\downarrow\uparrow \omeg= \omeg_d=\qquad 0  \quad 1 \quad 2 \quad 3 \quad \cdots.$$
This works because the only isomorphisms in $\omeg$ are identity morphisms. 
There is the obvious inclusion $\omeg_d \lra \omeg$ which comes from the forgetful functor from the iso-coslice category. 

We will also need the category of natural numbers with isomorphisms between each number.

The group of integers comes in many forms.
We can form the coeqaulizer 
$$ \xymatrix{ \so \ar@<-.1in>[r] \ar@<.1in>[r] & \omeg \coprod \omeg^{op} \ar[r]& {\mathbf \zeta}'}$$
where the two maps point to the $0 \in \omeg$ and the $0 \in \omeg^{op}$. This category looks like this  
$$\xymatrix{{\mathbf\zeta}' = &\cdots  &-3 \ar[l] &-2 \ar[l] &-1 \ar[l] &0\ar[l] \ar[r] & 1\ar[r]  &2\ar[r]& 3 \ar[r]& \cdots }$$
and is not very useful. A better category is obtained from the coequalizer 
$$ \xymatrix{ \so \ar@<-.1in>[r]_t \ar@<.1in>[r]^s & \sto \ar[r]^{\rho}& \overline\Z}$$
This gives us the group of integers as a one object category.  Taking the slice category gives us 
$$\xymatrix{{\mathbf \zeta}_i= \ast/ \overline{\Z}=& \cdots \ar[r]^{\sim}& -3 \ar[r]^{\sim}& -2 \ar[r]^{\sim} &-1 \ar[r]^{\sim} & 0 \ar[r]^{\sim} & 1 \ar[r]^{\sim}& 2 \ar[r]^{\sim} & 3 \ar[r]^{\sim}  & \cdots.  
}$$
While the isomorphisms between the objects of this category are important and we could work with this category, classical computability and complexity theory works with the natural numbers and does not place any importance on negative numbers. We can collapse the negative numbers to $0$ as follows:
$$\xymatrix{\omeg \ar@<-.1in>[r]_{inc} \ar@<.1in>[r]^c & {\mathbf \zeta}_i \ar[r]& \omeg_i}$$
where the $c$ map is a constant at $0$ and the $inc$ map is the inclusion of the natural number into the integers. This coequalizer essentially collapses the nonegitive integers to zero and leaves the rest alone. The category 
$$\xymatrix{\omeg_i= & 0 \ar[r]^{\sim} & 1 \ar[r]^{\sim}& 2 \ar[r]^{\sim} & 3 \ar[r]^{\sim}  & \cdots.}$$ is the (negative) natural numbers with unique isomorphisms between them. Since the nonpositive natural numbers are isomorphic to the nonnegative natural numbers, it does not matter which we use. $\omeg_i$ will be one of the major players in the tale we shall tell.

Each of these three categories ( $\omeg, \omeg_i$ and $\omeg_d$ ) have advantages and disadvantages. The $\omeg_d$ and $\omeg_i$ do not have interesting limits or colimits. In contrast $\omeg$ is a good target category only for a nondecreasing function.  $\omeg_i$ is good as a target category for arbitrary functions. The three categories 
$\omeg_d \hookrightarrow \omeg \hookrightarrow \omeg_i$
and the way we go from one to the other will be important.

We will also be interested in operations on these ordered categories. 
The successor operation on $\omeg$ is defined by first constructing the successor operation on the monoid of natural numbers:
$$\xymatrix{succ:\N\ar[r]^{\sim}&\N \times \bf 1 \ar[r]^{Id \times s}& \N \times \bf 2 \ar[r]^{Id \times \rho}& \N \times \N \ar[rr]^{MapComp}&& \N }$$
where $s$ is the source functor, $\rho$ takes the single arrow in $\bf 2$ to the first natural number, and \verb"MapComp" is the operation in the monoid.  
This functor works as it should: $succ(n)=n+1.$ 
With this, we can create the successor function on the linear order. Remembering that a slice category is a special type of comma category and for three functors $L:\A \lra \C$, $R:\B \lra \C$ and $H:\C \lra \D$ there is an induced functor of comma categories:
$$H_*: (L\downarrow R) \lra ((H L )\downarrow(H  R))$$
which takes $f:L(a) \lra R(b)$ to $H(f):H(L(a)) \lra H(R(b))$.
For the case of the total order of $\ast / \N$ this becomes
$$succ_*:\omeg \lra \omeg$$ 
which works as it is supposed to. 
There is a similar operation that can easily be defined on $\zeta_i$ and $\omeg_i$. When there is no ambiguity around, we refer to all the successor functors as $succ$. 
 
A number, $n \in N$, will correspond to a functor $P_n:\so \lra \omeg$ such that $P(0) = n.$ We can make similar definitions for 
$\zeta_i$ and $\omeg_i$. If $P$ is pointing to $n$, then $succ \circ P: \so \lra \omeg \lra \omeg$ points to the successor of $n$.

Finally we are ready to deal with computable functions.
\begin{defi}A total function $f: N^k \lra N$ is {\it constructible} if there is a Sammy program that describes a functor $F:\omeg_i^k \lra \omeg_i$ such that 
for all $x_1, x_2, x_3, \ldots, x_k$ in $N^k$ the following square commutes
$$\xymatrix{{\bf 1}\times {\bf 1}\times \cdots \times {\bf 1} \ar[rr]^{P_{x1} \times \cdots \times P_{xk}} && \omeg_i \times \omeg_i \times \cdots \times \omeg_i \ar[dd]^F
\\ \\ 
{\bf 1}\ar[uu]^{\sim}\ar[rr]_{P_{f(x_1,x_2, \ldots x_k)}}&&\omeg_i. 
}$$
\end{defi} 

\begin{teo}  Any totally computable function of natural numbers is constructible .
\end{teo}

\proof  First we show that the initial functions are constructible:
\begin{itemize}
\item The zero function $z: N \lra N$ defined as $z(n)=0$ is obtained as the composition of the unique function from $\omeg_i$ to $\so$ and the pointer $P_0:\so \lra \omeg_i$. That is, $$z=P_0 \circ !_{\omeg_i}: \omeg_i \lra \so \lra \omeg_i.$$ 
\item The projection functions are simply categorical projections functors $\pi^k_j: \omeg_i^k \lra \omeg_i$ for $1 \leq j \leq k$ which are induced by the product. 
\item The successor was already shown to be constructible.  
\end{itemize}  We also must show that constructibility is preserved under the operations of composition, recursion and the $\mu$-minimization operator. 
\begin{itemize}
\item Composition. If $f$ is constructible by the functor $F$ and g is constructible by the functor $G$, then function $g \circ f$ will be described by a Sammy program that describes $F$ and then $G$ and then gives the composition of the functors. We have to erase the $\verb"Return"$ statement at the end of the program for $F$. More sophisticated versions of composition are also easily done. 
\item Recursion. If $h$ is defined via recursion from $f$ and $g$ as 
$$h(\overline{x},0)=f(\overline{x}) \qquad \mbox{ and }\qquad h(\overline{x},n+1)=g( h(\overline{x},n))$$
and if $f$ and $g$ are constructible by $F$ and $G$, then $h$ is constructible by $H$.

We shall construct $H$ by using a (parameterized) natural number object. We remind the reader of  the basics of natural number objects. 
A natural number object in a category $\C$ is an object $N$ and two maps $z\colon 1 \lra N$ (to be thought of as using the terminal object 1 to chose the zero in $N$) and $s\colon N \lra N$ (to be thought of as the successor function) which satisfies the following universal property: If there is any object $A$ in the category with two maps $f \colon 1 \lra A$ and $g\colon A \lra A$ then there is a unique map $h\colon N \lra A$ such that both parts of the following diagram commutes
\be 
\xymatrix{1 \ar[rr]^z\ar[rrdd]_{f}&& N\ar[rr]^s\ar[dd]_h&& N\ar[dd]^h\\
\\
&& A \ar[rr]_{g}&&A. }
\ee 
Given a natural number object in a cartesian closed category (like $\Cat$), it is easy to construct a ``stable'' or ``parametrized'' natural number object (see section 5.5 of  \cite{CTforCS} or section I.9 of  \cite{LS}). This is a natural number object such that for any object $B$ and for any map $f\colon B \lra A$ and $g\colon A \lra A$, there is a unique $h\colon B\times N \lra A$ such that the following diagram commutes
\be 
\xymatrix{B \ar[rr]^{\langle Id_B, z \rangle }\ar[rrdd]_{f}&& B\times N\ar[rr]^{Id_B\times s}\ar[dd]_h&& B \times N\ar[dd]^h\\
\\
&& A \ar[rr]_{g}&&A. }
\ee 

We will show how to construct a natural number object in $\Cat$, i.e., the initial object in the category $\Cat^{\so \lra X \lra X}$. (Another way of dealing with this is saying that a natural number object is the initial algebra for the $X \mapsto (X + \so)$ endofunctor. We will not go this way.)   
First let us construct $\Cat^{\so \lra X \lra X}$. We do this with two equalizers. Start by easily constructing the category $\cdot \lra \cdot \lra \cdot$. Now take the equalizer 
$$\xymatrix{\Cat^{\cdot \lra X \lra X}\ar[rr] && \Cat^{\cdot \lra \cdot \lra \cdot} \ar@<1ex>[rr]^{ev_2}\ar@<-1ex>[rr]_{ev_3}&&\Cat}$$ 
where $ev_2$ and $ev_3$ are the evaluations at the second and third $\cdot$ of the diagram. Now find an even smaller subcategory of this functor category by taking the following equalizer 
 $$\xymatrix{\Cat^{\so \lra X \lra X}\ar[rr] && \Cat^{\cdot \lra X \lra X} \ar@<1ex>[rr]^{ev_1}\ar@<-1ex>[rr]_{Const_{\so}}&&\Cat}$$ 
 where $ev_1$ is evaluation at first $\cdot$ and $Const_{\so}$ is the constant functor that picks out the terminal category $\so$. To find the initial object in this category we simply take the following Kan extension:
$$ \xymatrix{\so \ar[rrrr] &&&& \Cat^{\so \lra X \lra X}\\
 \\
 &&{\bf 0}.\ar[lluu]_\!\ar[rruu]^\!}$$
 We claim that the initial object in this category is 
 $$\xymatrix{\so \ar[rr]^{P_0}&& \omeg_i \ar[rr]^s&& \omeg_i}.$$ 
 Let us use this paramaterized natural number object with the object $B=\omeg_i^n$.
Given any $F\colon \omeg_i^n \lra \omeg_i$ and $G\colon \omeg_i \lra \omeg_i$, we have an object in the category $\Cat^{\so \lra X \lra X}$. The morphism $H\colon \omeg_i^n \times \omeg_i \lra \omeg_i$ is constructed from the induced map from the Kan extension (the morphism from an initial object of a category).

\item $\mu$-minimization. Let $f:N^k \times N \lra N$ be a total function. The $\mu$-minimization of $f$ is a function $g:N^k \lra N$ that is defined for an $\overline{x}\in N^k$ as the smallest $y$ such that $f(\overline{x},y)=0$. We assume that $f$ has the following property: for all $\overline{x} \in N^k$ there exists a $y\in N$ such that $f(\overline{x}, y)=0$. This makes $g$  a total function.  If $f$ is constructible by $F:\omeg_i^k \times \omeg_i \lra \omeg_i$ then $g$ will be constructible by $G:\omeg_i^k \lra \omeg_i$ where $G$ is defined by the composition of the double  lines in the following diagram:
$$\xymatrix{\omeg_i^k \sim \omeg_i^k \times \so \ar@{=>}[r]^L\ar[ddr]_{Id\times 0} &{\mathbf\alpha} \ar@{=>}[dd]_{H}\ar[rr]^{!}&& \so\ar[dd]^{P_0}
\\ \\
&\omeg_i^k \times \omeg\ar@{=>}[r]^{Id \times inc} &\omeg_i^k \times \omeg_i\ar[r]^F\ar@{=>}[dd]_{\pi}& \omeg_i \\ \\
&&\omeg_i}$$
In detail, first precompose $F$ to get the bottom line $\omeg_i^k \times \omeg \lra \omeg_i^k \times \omeg_i \lra \omeg_i$. Also take the functor $P_0:\so \lra \omeg_i$. Now take the pullback to get ${\mathbf\alpha}$.  
Notice that ${\mathbf\alpha}$ is the full subcategory of $\omeg_i\times \omeg$ of all the $y$ such that $F(\overline{x},y)=0$. We would like the smallest one of these. Both $\omeg_i^k$ and ${\mathbf\alpha}$ are completely contactable groupoids and so to find the smallest one use a left Kan lifting $L$ of $Id \times 0$ that picks out $(\overline{x},0)$ and the inclusion of $H$.  
$L$ will pick out find the least of all $y$ in ${\mathbf\alpha}$. Notice the triangle on the left does not commute. Let 
$$G= \pi \circ  inc \circ  H\circ L : \omeg_i^k \lra {\mathbf\alpha} \lra \omeg_i^k \times \omeg \lra \omeg_i^k \times \omeg_i \lra \omeg_i.$$ 
\end{itemize}
$\Box$

\vspace{.2in}
While we can use computable functions to discuss computability, we can also use Turing machines. Turing machines are important for dealing with complexity. The usual definition of the time complexity of an algorithm is the amount of time clicks a Turing machine uses in order to complete its mission. If we are going to compare category theory with complexity, we will need to show how category theory can mimic Turing machines. 

Our Turing machine tape will be the category 
$$\omeg =\qquad  0 \lra 1 \lra 2 \lra 3 \lra \cdots. $$ 
The head of a the Turing machine will be a functor $P_i:\so \lra \omeg$ pointing to the $i$th position. To move the pointer to the right, we use the successor function: 
$$P_{i+1}=succ \circ P_i: \bf 1\lra\omeg \lra \omeg.$$
Similarly, if $P_i:\so \lra \omeg$ points to the $i$th position, then the predecessor  
$P_{i-1}: \so\lra\omeg$ is constructed using the Kan lifting:
\be\xymatrix{\so \ar[rr]^{P_{i-1}}\ar[rd]_{P_i}&& \omeg\ar[ld]^{succ}\\
&\omeg.}\ee
This will permit us to move to the left and right on the Turing machine tape. 

If $P_i:\bf 1 \lra \omeg$ points to the $i$th position on the tape, then the comma category constructions gives us truncated tapes:
$$\xymatrix{(\omeg \downarrow P_i)=& 0\ar[r] & 1 \ar[r] & 2 \ar[r] & 3 \ar[r] & \cdots \ar[r]&i-1\ar[r]&i }$$
and 
$$\xymatrix{ (P_i \downarrow\omeg)=& i\ar[r] & i+1 \ar[r] & i+2 \ar[r]& i+3 \ar[r] & i+4 \ar[r]  & \cdots . }$$
There are canonical inclusions of these subcategories into $\omeg$. These constructions will permit us to partition the tape and manipulate parts of it.

The tape alphabet will be $\{ 0, 1, \Box \}$ where the $\Box$ will correspond to the blank. Using coequalizers with copies of ${\mathbf 2}$ and $\overline{\mathbf 2}$, it is easy to construct the following two categories and the inclusion of the first into the second
$$
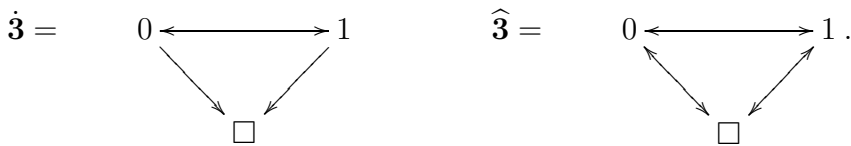
\xymatrix{\dot{\bf{3}}= & 0  \ar@{<->}[rr]\ar[dr]&&  1 \ar[dl]\\ 
&&\Box}
\qquad \qquad\xymatrix{\widehat{\bf{3}} = & 0  \ar@{<->}[rr]\ar@{<->}[dr]&&  1 \ar@{<->}[dl]\\ 
&&\Box}.$$
A functor $F:\omeg \lra \widehat{\bf{3}}$ assigns a value in the tape alphabet to every position on the tape.
$F$ is a string in $\{ 0, 1 \}$ if it starts at $0$ or $1$ and factors as follows:
$$
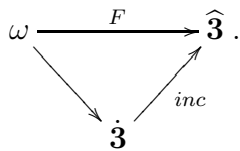
\xymatrix{ \omeg  \ar[rr]^F \ar[dr]&&  \widehat{\bf{3}} \\ 
&\dot{\bf{3}}\ar[ur]_{inc}}.$$
(We can tell if a functor factors by taking a Kan lifting and seeing if the associated $\alpha$ is the identity natural transformation.) 

Now to deal with the contents of the tape.
 There are three distinguished functors $0:\bf 1 \lra \widehat{\bf{3}}$, 
$1:\bf 1 \lra \widehat{\bf{3}}$, and 
$\Box:\bf 1 \lra \widehat{\bf{3}}$. The contents of $i$th position of a Turing machine tape is
$$F \circ P_i:\bf 1 \lra \omeg \lra \widehat{\bf{3}}.$$
If we would like to do some procedure when the $i$th position of the tape has a 1 in it, we can use this in a Sammy program:
$$\verb"If " F o P_i \verb" == " 1 \verb" goto L".$$
We put all these parts together to make a Turing machine. First we use a Sammy program to make $\omeg$, $\dot{\bf{3}}$ and $ \widehat{\bf{3}}$. There will also be a variable $q_x$ which will point to some element of the discrete category $Q=\{0, 1, 2, ,\ldots, |Q|\}$ that will correspond to the state of the Turing machine. The input to the Turing machine will be given as a functor $F:N \lra \widehat{\bf{3}}$. There will be a pointer $P_0:\bf 1 \lra \omeg$ that begins by pointing at the zeroth position of the tape. What remains is to show how to translate a transition rule of a Turing machine into a Sammy program.

Let us say that the Turing machine rule says
$$\delta(q_{35},a)=(q_{52},b,D)$$
where $a,b\in \{ 0,1,\Box\}$ and $D \in \{ R, L \}$. This rule will be done when the conditional

{\verb If $F\circ P_i\verb" == "a$ AND $q_x\verb" == " 35$ \verb" goto L"

\noindent is satisfied. If this condition is met, then the following algorithm in Sammy will be executed:

$$
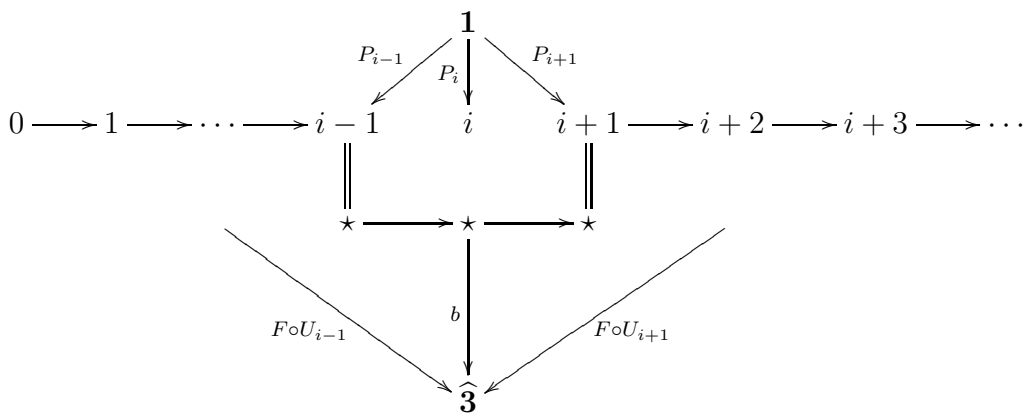
\xymatrix{&&&&\bf 1\ar[ld]_{P_{i-1}} \ar[d]_{P_{i}}\ar[rd]^{P_{i+1}}\\
0 \ar[r]&1\ar[r]& \cdots \ar[r]&i-1\ar@{=}[d]& i&i+1 \ar@{=}[d]\ar[r]&i+2\ar[r]&i+3 \ar[r]&\cdots\\
&& \ar[rrdd]_{F\circ U_{i-1}}&\star\ar[r]&\star\ar[dd]_b\ar[r]&\star&\ar[lldd]^{F\circ U_{i+1}}\\
\\
&&&&\widehat{\bf{3}}
}$$

\begin{enumerate} 
\item Form $P_{i-1}$, $(\omeg\downarrow P_{i-1})$, and forgetful functor  $U_{i-1}:(\omeg\downarrow P_{i-1})\lra\omeg$. This is the left side of the tape.
\item Form $P_{i+1}$, $( P_{i+1}\downarrow\omeg)$, and forgetful functor  $U_{i+1}:(P_{i+1}\downarrow \omeg)\lra\omeg$. This is the right side of the tape.
\item Use composition to form $F \circ U_{i-1}:(\omeg\downarrow  P_{i-1}) \lra \omeg \lra \widehat{\bf{3}}$ and $F \circ U_{i+1}:(P_{i+1}\downarrow  \omeg) \lra \omeg \lra \widehat{\bf{3}}$. These two sides retain their old values.
\item Consider $b:1 \lra \widehat{\bf{3}}$. This is the new value at point $i$ of the tape.
\item Use coeqalizers to attach $(\omeg\downarrow P_{i-1})$ with $\bf 2$ and $\bf 2$ to $\bf 1$. Connect the left side of the tape to position $i$.
\item Use coqualizers to attach $\bf 1$ to $\bf 2$ and $\bf 2$ to $( P_{i+1}\downarrow\omeg )$. Connect the right side of the tape to position $i$.
\item Since all three of these parts have a map into $\widehat{\bf{3}}$, there is an induced functor from the combination of all three parts into $\widehat{\bf{3}}$. This is the new tape and values.
\item Set $q_x=52$.
\item If $D=R$ then $P_i$ is set to its successor, else $P_i$ is set to its predecessor. 
\end{enumerate}

Conclusion: We showed that for every rule of the Turing machine there is a set amount of steps of a Sammy program. Hence our programming language can do whatever Turing machines can do. For a string $s$, there is a functor
$F_s:\omeg \lra \dot{\bf{3}}$ and our programming language can manipulate the Turing machine tape in the same way that the Turing machine would manipulate the tape.
 What this means is that for the computational complexity to create the functor $F_s$ is the same as the computational complexity for a Turing machine to create $s$.  
The size of the Sammy program is, up to a multiplicative constant, the same as the number of rules in the Turing machine. In particular, if we consider the smallest programs, we get 
\begin{teo}
For every string $s$, $K_{Sammy}(F_s)= O(K_{Classical}(s)).$ 
\end{teo}
In a sense, this says that our Kolmogorov complexity is a generalization of classical 
Kolmogorov complexity. 

\vspace{.4in} 

We will need to look at the Kolmogorov complexity for numbers. In classical Kolmogorov complexity, for any natural number $n$,  $K(n) \le log_2n$. For categories, the analogous theorem is as follows:
\begin{teo}For any natural number $n$, the functor $P_n: \so \lra \omeg$ can be constructed by Sammy in $O(log_2n)$ operations. That is, $K_{Sammy}(P_n)\le O(log_2 n).$ 
\end{teo}

\proof We must compose with the successor function $n$ times on $P_0$ in order to get $P_n$.  Basically, the idea is that one can look at the binary representation of $n$ and write a program based on that. For example 
727 can be written in binary as 1011010111. We can express this number as 
$$(((((((((1 \times 2+ 0)  \times 2+ 1) \times 2+ 1) \times 2+ 0) \times 2+ 1) \times 2+ 0)\times 2+1)\times 2+1)\times 2+1). $$ Similarly, we can start with $P_0$ and $P_x=P_1$ and then look at each digit of the binary expression of $n$ and either \newline (a) double the length between $P_0$ and $P_x$ or 
\newline (b) double the length between $P_0$ and $P_x$ and then compose with the successor function one more time. 

For example the following algorithm will perform option (b):
\begin{enumerate}
\item $P_t=P_0$
\item $P_{end} = P_x$
\item $\verb"If" P_t\verb" == " P_{end} \verb" goto"$ 7 
\item $P_x =succ \circ P_x$
\item $P_t = succ \circ P_t$
\item goto 3.
\item $P_x= succ \circ P_x$
\item $\verb"Return("P_x\verb")"$
\end{enumerate}
 For option (a),  simply leave out line 7. 

For a given $n$, a Sammy program that will either consist of a sequence of  option (a) or option (b) will perform the required task.  The length of this program will be 6 or 7 times as long as $log_2~n$. 
\endproof

In the above, for every natural number $n$ there will be a different Sammy program. One can, instead, think of having the number $n$  inputed as a binary string. Such a string will be a functor $F: (\omeg \downarrow log n) \lra \sto$.  
A constant sized Sammy program can then produce the functor $P_n$. Basically the Sammy program reads the input by looking at the functor $F$ deciding at each point to do option (a) or option (b). The pseudo-code will look like this
\begin{enumerate}
\item $P_i=P_0$
\item $\verb"If"  F \circ P_i \verb" == " 0$, then perform option (a) else perform option (b)
\item  $\verb"If" P_i\verb" == " P_{log n} \verb" goto"$ 6
\item $P_i = succ \circ P_i$
\item goto 2.
\item $P_n= P_x$
\item $\verb"Return"(P_n)$

\end{enumerate}

We have just proved that 
$$K_{Sammy}( P_n:{\bf 1} \lra \omeg \quad |\quad F: (\omeg \downarrow log n) \lra \sto)\qquad = \qquad O(1)$$
where $F$ describes $n$ in binary.

\vspace{.2in}

In the future we will need to know when a category is discrete. $\A$ is a discrete when the map induced by $s:{\bf 1} \lra {\bf2}$, 
$$\A^{\bf 2} \lra \A^{\bf 1}$$ 
is an isomorphism. This basically says that the only morphisms in $\A$ are identities. 

We will also need to know when a category is connected. $\A$ is connected when $$({\bf 1}+{\bf 1})^{\A}\sim {\bf 1}+ {\bf 1}.$$
That is, a category is connected when there are exactly two morphisms from the category to the discrete category with two objects.

\vspace{.2in}
The most celebrated limitation of theoretical computer science is that no computer can solve the halting problem. That is, if $\phi_y$ is the function that corresponds to the $y$th computer program and $x$ is an input, there is no computer program that can tell if $ \phi_y(x)$ will halt or go into an infinite loop. The categorical analog for this problem is to construct the functor $Halt:\omeg_i \times \omeg_i \lra \sto$
$$  Halt(x,y)= \left \{
\begin{array}{r@{\quad : \quad}l}
1 & \mbox{if } \phi_y(x)\downarrow\\
0 & \mbox{if } \phi_y(x)\uparrow \end{array}\right.$$

Whereas no regular computer can describe the halting predicate, categories can. 
\begin{teo}
Halt is a constructible functor. 
\end{teo}

\proof It is known that the predicate $Halt'$ defined as  
$$  Halt'(x,y,t)= \left \{
\begin{array}{r@{\quad : \quad}l}
1 & \mbox{if } \phi_y(x)\downarrow \mbox{ within } t \mbox{ steps} \\
0 & \mbox{if } \phi_y(x)\uparrow \mbox{ within } t \mbox{ steps} \end{array}\right.$$
is total and computable and hence constructible in Sammy.

Notice that for a given $x$ and $y$, once $Halt'(x,y,t)$  becomes true,  it remains true for any $t'\ge t$.
This means that the constructible functor $Halt''=Halt'\circ (inc \times inc \times Id):\omeg_d \times \omeg_d \times \omeg_i \lra \omeg_i \times \omeg_i \times \omeg_i \lra \sto$ factors through $\st$ as in 
$$\xymatrix{\omeg_d \times \omeg_d \times \omeg_i \ar[rr]^ {Halt''}\ar[dr]_{Halt_*}&&\sto\\
&\st\ar[ur]_{inc}
}$$
Another way to say this is that the Kan lifting of $Halt''$ is $Halt_*$.

$Halt_*$ can be used in the following right Kan extension to construct the Halting functor:
$$\xymatrix{\omeg_i\times \omeg_i \ar@{-->}[rr]^{Halt} && {\bf 2} \\ &\omeg_d\times \omeg_d\times \omeg_i. \ar[lu]^{(inc \pi^3_1) \times (inc \pi^3_2)}\ar[ru]_{Halt_*}}$$
For a given $x$ and $y$, the halt functor has the value 
$$Colim(0 \lra 0 \lra   0 \lra  0 \lra  0 \lra  0 \lra  0 \lra \cdots) = 0$$
if program $y$ does not halt on input $x$. In contrast, if program $y$ on input $x$ does eventually halt, then the value of the functor is 
$$Colim( 0 \lra  0 \lra  0 \lra  0 \lra  0 \lra  0 \lra \cdots  0 \lra  0 \lra  1 \lra  1 \lra  1 \lra  1 \lra \cdots )=1.$$ 
We will use these idea again in Section 5.

\endproof

Essentially this follows from the intuition that $$Halt(x,y)= Colimit_tHalt'(x,y,t).$$ This should however not be surprising since categories can do infinitary operations. However this says something about implementations of Sammy. To reiterate what was said above: since we know from the Turing's halting theorem and the Church-Turing thesis that the halting problem cannot be solved by any computer, we know that the Sammy programming language can not be totally 
implemented on any computer.

\section{Nonconstructible Categorical Structures}

Not every categorical structure is constructible. A simple counting argument can be used to see this. A functor $\omeg \lra \sto$ corresponds to a real number. Therefore there exists uncountably infinite functors from $\omeg$ to $\sto$. However, every Sammy program can be encoded as a natural number. This means that there are only countably infinite constructable functors from $\omeg$ to $\sto$. The vast vast majority of structures are not constructible.  

A concrete example of a functor that is not constructible is $K_{Sammy}$. It is well known $K_{Classical}: Strings \lra N$ is not a computable function. First let us be careful about the definition of $K_{Sammy}$. It is a functor that assigns to every category, functor, and natural transformation a natural number. We might as well assume that it only assigns natural transformations since identity natural transformations are simply functors and identity functors are simply categories. Let us think of $\Cat$ as the discrete category of natural transformation. We are going to forget the (two) composition structures on $\Cat$ because $K_{Sammy}$ does not behave well with respect to maps. Another problem is that not all elements of $\Cat$ are constructible. If $\X$ is not constructible we shall let $K_{Sammy}(\X)= \infty.$ So we have a functor $K_{Sammy}:\Cat \lra \overline{\omeg}$ where $\overline{\omeg} = \omeg \cup \infty$ is the completed natural numbers. 
\begin{teo}
$K_{Sammy}:\Cat \lra \overline{\omeg}$ is not constructible.
\end{teo}  

\proof The proof is a self-referential proof by contradiction argument similar to the Berry paradox. Assume (wrongly) that $K=K_{\Sammy}$ is, in fact, constructible, then there is a shortest program that creates $K$. So we can ask for the finite value of $K(K)$ (this is the core of self reference!). Let $K(K)=c$. Let $n$ be a natural number and let $P_n:\bf 1 \lra \overline{\omeg}$ be a functor such that $P_n(0)=n$. Now use $K$ and and $P_n$ to construct the following pullback:
$$\xymatrix{\Cat_n \ar@{^{(}->}[rr]\ar[dd]&& \Cat\ar[dd]^K 
\\ \\
 (P_n\downarrow \overline{\omeg}) \ar@{^{(}->}[rr]&&\overline{\omeg} .
}$$
$ (P_n \downarrow\overline{\omeg)}$ is the sub-total order of the completed natural numbers that start at $n$. $\Cat_n$ is the discrete set of natural transformations whose shortest program is greater than or equal to $n$ lines. For any $n$, to construct  $\Cat_n$ one needs 
\begin{enumerate}
\item $c$ lines of code to construct $K$
\item $log n$ lines to construct $P_n$
\item and a few lines to construct the pullback. Call this small number $pb$.
\end{enumerate} 
That is, $\Cat_n$ can be constructed in $c+log n + pb$ lines or $K(\Cat_n)\leq c + log~ n+pb$.  
Choose an $n$ such that $n >> c + log~ n+ pb$. Then $\Cat_n$ contains objects that require $n$ or more lines of code (or are unconstructible) while we just described $\Cat_n$ in $c + log~ n+pb$ lines of code.We assume that $\Cat_n$ has the axiom of choice.  This is like a Berry sentence. Contradiction! The only thing assumed is that $K$ was constructible. It is not constructible. 
\endproof

What else is not constructible? Consider the functor that determines if a categorical structure is constructible or not. The functor $Cons: \Cat \lra \sto$ which accepts a natural transformation and tells if that natural transformation is constructible or not. While, at the moment, I cannot prove it, I conjecture that $Cons$ is not constructible.

In the last section we showed that one can write a Sammy program to solve the halting problem for classical Turing machines. What about the halting problem for Sammy programs? There are several issues here to deal with. First off, We have seen that some programs will go on for $\omega$ steps and then stop. This is similar to the infinite time Turing machines. Do we say that such problems halt?  Another point is that the halting problem is not about a program halting, rather it is about a program {\it and an input to the program.} The analogous categorical question would ask about a functor $F: \A \lra \B$ and a input functor $I: \star \lra \A$. With these two functors we can ask if 
$F \circ I$ is defined or is constructible.
 
 Throughout this paper, the structures that we dealt with were totally defined. Category theory usually deals with totally defined structures. Mathematicians do not talk about ``partially defined cohomology functors.'' Physicists generally do not talk of ``a partially defined Lagrangian.'' In fact, essentially the only time that category theorists talk about partial functors is to mimic something from computability theory.

So even though this is orthogonal to our main purpose, we will spend a little time looking at partially defined functors. In the spirit of \cite{DipHel} one can define a partial functor $F: \C \longrightarrow \D$ as a pair$(inc_F, F_0)$ where $inc_F$ is an inclusion of a subcategory $inc_F: \C_F \longrightarrow \C$ and $F_0:\C_0 \longrightarrow \D$ is a functor from that subcategory to $\D$. Composition of $F=(inc_F, F_0): \C \longrightarrow \D$ and $G=(inc_G, G_0):\D \longrightarrow \E$ is given by the following pullback:
$$ 
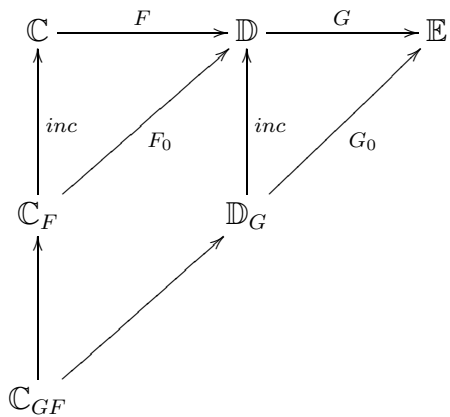
\xymatrix{ 
\C \ar[rr]^F&& \D\ar[rr]^G && \E \\  \\
\C_F\ar[uu]_{inc}\ar[uurr]_{F_0}&&\D_G \ar[uu]_{inc}\ar[uurr]_{G_0}\\ \\
\C_{GF}\ar[uu]\ar[uurr]} 
$$
The problem with this is that composition is only associative up to a natural transformation. This takes us out of the domain of category theory and into the the field of bicategories. One would have to formulate the notion of a natural transformation between two partial functors and then go on to describe Kan extensions and Kan liftings of partial functors. The Sammy programing language would have to be modified to Sammy' which describes  categories, partial functors and partial natural transformations between them. In this setting, there will be many analogs to the theorems in classical computability theory.
Sammy' programs are very simple objects. Sammy' only has a finite number of operations and each Sammy' program consists of a finite number of lines. So every Sammy' program can be encoded as a finite number and hence a functor $P:\so \lra \omeg$. From this it follows that we can enumerate all Sammy' programs. 

We are only interested in Sammy' programs whose returned value is a partial functor $\omeg_i \lra \sto$.
Formally, let $\phi^S_n:\omeg_i \lra \sto$ be the partial functor described by the $n$th Sammy' program. If the $n$th program 
describes a partial functor from a different source or target category, then let it be undefined. Let $x$ and $y$ both be thought of as natural numbers and as functors $\so \lra \omeg$.
Consider the functor $Halt^S:\omeg_i \times \omeg_i \lra \sto$ that is defined as 
$$  Halt^S(x,y)= \left \{
\begin{array}{r@{\quad : \quad}l}
1 & \mbox{if } \phi^S_y(x)\downarrow\\
0 & \mbox{if } \phi^S_y(x)\uparrow \end{array}\right.$$

\begin{teo}
The total functor $Halt^S$ is not a constructible functor.
\end{teo}   
\proof The proof will follow the usual self-referential ideas found in the classical proof that the halting problem is unsolvable (see e.g. \cite{Universal}.) Assume (wrongly) that $Halt^S:\omeg_i \times \omeg_i \lra \sto$ was a total constructible functor. We can compose this total functor with the following easily constructible  total functors $\Delta: \omeg_i \lra \omeg_i \times \omeg_i$ and $NOT:\sto \lra \sto$ such that $NOT(0)=1$ and $NOT(1)=0$ and get 
$$\xymatrix{&&\omeg_i \times \omeg_i \ar[rr]^{Halt^S}&&  \sto\ar[dd]^{NOT} 
\\ \\
\so \ar[rr]_{P_y}&&\omeg_i \ar[rr]_D\ar[uu]^{\Delta}&&\sto.
}$$
By composition we get the functor $D: \omeg_i \lra \sto$ (for diagonal). $D$ is defined so that $D(n)=1$ iff Sammy' program $n$ on input $n$ is not constructible. If $Halt^S$ was constructible than $D$ would be constructible and there would be a program number, say $y$, for this program. That is $D=\phi^S_y:\omeg_i \lra \sto$. Composing $D$ with $P_y:{\bf 1} \lra \omeg_i$ would make a functor 
$$D \circ P_y=\phi^S_y \circ P_y: {\bf 1} \lra \omeg_i \lra \overline{2}.$$
We then have
$$\phi^S_y(y)=1 \mbox{   iff   } D(y)=1 \mbox{   iff   } NOT (  Halt^S(y,y))=1 \mbox{   iff   } Halt^S(y,y)=0 \mbox{ iff }\phi^S_y(y)=0$$
 We conclude that $D$ cannot be constructible which means that  $Halt^S$ cannot be constructible. 
\endproof

\section{The Arithmetic and Analytic Hierarchies}
Logicians have defined several hierarchies of logical predicates. The simplest such hierarchies is called the arithmetic hierarchy. This is the class of predicates or relations on natural numbers that are either recursive or can be gotten from recursive relations from either complementation and/or projections. Complementation corresponds to the $\neg$ operation while projection corresponds to $\exists$. Noticing that $\forall \equiv \neg \exists \neg$ and that quantifiers can be brought to the beginning of a logical formula, we can write all predicates in the arithmetic hierarchy as a sequence of existential and universal quantifiers followed by a recursive relation. The classification is then based on whether the first quantifier is a existential or universal, and on the number of alternations between the types of quantifiers. A predicate that can be expressed as an existential quantifier followed by $n$ alternations between universal and existential quantifiers will be called $\sum^0_n$. A predicate that can be expressed with a universal quantifier followed by $n$ alternations will be called $\prod_n^0$.
 Our main source and our notation will follow chapters 14-16 of \cite {rogers}.


\begin{teo} Every predicate in the arithmetic hierarchy is constructible. 
\end{teo}

\proof The proof is by induction on the complexity of the predicate. 
\begin{itemize}
\item If $\phi$ is a recursive predicate of $n$ variables, then, since we can think of $\sto$ as a subcategory of $\omeg_i$, and  as we saw about such recursive functions, there is a constructible functor $\Phi:\omeg_i^n \lra \sto$ that mimics $\phi$.
\item If $\phi \equiv \neg \psi$, and $\psi$ is mimicked by the constructible $\Psi :\omeg_i^n \lra \sto$, then $\phi$ is mimicked by the composition of constructible functors $$\Phi=NOT \circ \Psi: \omeg_i^n \lra \sto \lra \sto$$ where $NOT$ is the easily constructed functor that such that $NOT(0) =1$ and $NOT(1)=0$.     
\item Let $\phi \equiv \exists x \psi$ where $\psi$ is $n+1$ variable predicate that is mimicked by $\Psi:\omeg_i^{n+1} \lra \sto$. We shall construct $\Phi:\omeg_i^n \lra \sto$ in four steps.
\begin{enumerate}
\item Change the source of $\Psi$ to be partially discrete 
$$\Psi'=\Psi \circ (inc^n \times Id): \omeg_d^n \times \omeg_i \lra \omeg_i^n \times \omeg_i \lra \sto.$$
We will need this so that step 3 will work out well.
\item Change the target of $\Psi'$. Since there are no nontrivial limits or colimits in $\sto$ we must change the target category to $\st$ using a Kan lifting:
$$\xymatrix{\omeg_d^{n}\times \omeg_i \ar[ddr]_{\Psi'} \ar[rr]^{\hat{\Psi}}&&\st\ar[ldd]^{inc}\\ \\&\sto}$$ 
\item Now we use $\hat{\Psi}$ in the following right Kan extension
$$\xymatrix{
\omeg_i^n \ar[rr]^{\hat{\Phi}}  && \st
\\
\\
&\omeg_d^{n}\times \omeg_i \ar[luu]^{inc^n\circ \pi^n} \ar[ruu]_{\hat{\Psi}}
}$$
\item To get back a functor whose range is $\sto$ we simply compose $\hat{\Phi}$ with the obvious inclusion of $\st$ to $\sto$:
$$\Phi: \omeg_i^n \lra \st \lra \sto.$$
\end{enumerate}
\end{itemize}
We do these inductive steps for each complimentation and projection. In the end, we get a functor $\so \lra \sto$ which tells the truth or falsity of the predicate. \endproof

\vspace{.2in}

While we showed that categories can deal with all of the arithmetic hierarchy, we left open the question of the analytic hierarchy? Categories cannot solve everything. As we saw, they cannot construct the $K_{Sammy}$ functor. So what exactly is the power of categories?

In a sense, set theory can describe all the effective hierarchies and hence is less interesting. In contrast, category theory is limited. Some problems they can solve and some they cannot. This makes the study of the effective hierarchies from the categorical standpoint important. 

An interesting problems in the analytic hierarchy is to determine  if a given real number is a rational number. It is interesting to look at the problem from a categorical point of view. Think of  a real number as a functor $r: \omeg \lra \sto$. In order to describe a rational number we need the notion of a {\it lollipop}.
For every two nonnegative integers $m\leq n$,  we shall call the following partially ordered category $L_{m,n}$
$$\xymatrix{0\ar[r]&1\ar[r]&2\ar[r]&\cdots\ar[r]&m-1\ar[r]&m\ar[r]&m+1\ar[r]&m+2\ar[d]\\
&&&&&&&\vdots\ar[d]\\
&&&&&n-1\ar[uu]&n-2\ar[l]&n-3\ar[l]
}$$ 
There is always a canonical projection $\omeg \lra L_{m,n}$. It is not hard to see that $r$ is rational number if there is some $m$ and $n$ such that $r$ factors as 
$$\xymatrix{\omeg \ar[rr]^r \ar[rd]&& \sto \\& L_{m,n}\ar[ru]}.$$
One can look at the category of lollipops with $\omeg$ as the initial object and ${\bf 1}=L_{0,0}$ as the terminal object. However it is not clear how one constructs this category of lollipops or sees if an arbitrary real number factors through any of them. Much work remains. 

\section{Future Directions}

We see this paper as just the beginning of a larger project to understand the computability and complexity of categorical structures. There are many more subjects that we would like to attack with the tools developed here. Here are a few of them.

\vspace{.2in} 

\noindent{\bf Complexity Measures.} We would like to better clarify what is constructible and what is not. We discussed the Kolmogorov complexity of categorical structures. We would also like to study different complexity measures on these structures. Rather than asking what is the shortest program that produces a categorical structure, we can ask how much time/space does a program take to create a certain structure. That is, what is the computational complexity of a structure. We can also ask how much time does it take for the shortest program to produce that structure (Bennett's logical depth). Another  measures of interest are Adelmann's time-bounded Kolmogorov complexity. All these measures induce hierarchies and classifications of categorical structures. 

\vspace{.2in}

\noindent {\bf Entropy.} There is a relationship between classical Kolmogorov complexity and Shannon's complexity theory. Classical Kolmogorov complexity measures the complexity of an individual string while Shannon's complexity measures the complexity of a source of strings, or a whole class of strings. Shannon's entropy function, $H$, measures how much rigidity there is in the source. Or to put it another way, how many different ways can the strings be swapped. The relationship says that if $X$ is a source of messages or a class of messages, than $H(X)$ is about the average of all the $K(x)$ where $x$ is a string that can be produced by $X$. Formally, 
$$H(X)\sim \sum_ {x_i\in X}p(x_i)K(x_i).$$  

We would like to formulate a notion of Shannon's complexity theory for categories. There should be a definition of entropy of a category. This should measure how rigid or flexible a categorical structure is. Let 
$\C$ be a category, then $Aut(\C)$ is the group of automorphism functors $F:\C \lra \C$.
Define the ``entropy'' (or ``Hartley entropy'') of $\C$ as 
$H(\C)= Log_2 |Aut(\C)|$
(perhaps this should be
$H(\C)= p Log_2 \frac{1}{|Aut(\C)|}$
where $p$ is some constant.)
There is a similar definition for the ``entropy'' of a functor $F:\C \lra \D$
as
$H(F)=Log_2|Aut(F)|$
where we mean the set of iso-natural transformations $F \lra F.$
We might look at the entropy of a particular object $c$ of $\C$ as the
entropy of the functor $c:\so \lra \C$ that ``picks'' an object $c \in \C$.
$H(c)=H(c:\so \lra \C)=Log_2|Aut(c)|.$
It would be interesting to see the relationship between such entropy measures for categories and functors with $K_{Sammy}(\C)$ and $K_{Sammy}(c: \so \lra \C).$ Just as there is a relationship between these measures for strings, there should be a relationship for categorical structures.

\vspace{.2in}

\noindent{\bf Categories With Structure.}  So far we have restricted to classical categories, functors, and natural transformations. But that is not the whole story. What about categories with more structure? For example, what can we say about a category that we know has all limits and colimits? What about enriched categories, 2-categories, n-categories, weak-n-categories, quasi-categories, categories with Quillen model structures (i.e., the ability to do homotopy theory) etc? These different structures have been applied in almost every area of mathematics, computer science and theoretical physics. For each of these classes of categories with structure, we are going to have to come up with another Sammy programing language that can deal with such structure. We are interested in what structures of these types can and cannot be constructed and what is their complexity.  For example when dealing with 2-categories, we might be interested in given the category $\B$ to construct the 2-category $Fib(\B)$ of fibrations over $\B$. We also would like to see its relationship with $[\B^{op}, \Cat]$. 

\vspace{.2in}

\noindent{\bf Classical Computational Complexity Theory.} There is an intuition that the difference between P and NP is that the solution space for P is somehow structured while the solution space for NP is unstructured. An algorithm in P uses the structure of the solution space to find the correct solution efficiently. In contrast, the only known way of solving an NP problem is to perform a brute-force search through the unstructured solution space. We believe that it should be possible to formalize this 
intuition with our Kolmogorov complexity of categories. What does it mean for a solution space to have a workable structure? Most of the solution spaces for NP problems do have some type of structure. They usually form an exponential tree. Why is that not good enough? What about other complexity classes. How does PSPACE compare with NP? There must be some type of trade-off  between the size of the solution space and the amount of structure of the solution space. If the size is very large but there is a lot of structure, then the problem could still be effectively solved. In contrast, if the solution space is relatively small but there is little structure, the problem will be hard to solve. We hope to examine this notion of structured solution space with Kolmogorov complexity. 

\newpage
\section{Appendix: Syntax and Semantics of Sammy}
In order to be more exact we give the formal syntax and semantics of Sammy.

\vspace{.3in}

{\bf \qquad \qquad \qquad The EBNF Syntax for Sammy}

\begin{tabular}{rcl}

program &$\lra$&\{~~statement~~\}~~ return\\
statement  &$\lra$&  [ label ] (~~assignmentStmt    $|$   ifStmt  ) \\
label  &$\lra$&  (~~$\verb"A" ~~|~~ \verb"B" ~~|~~    \cdots ~~|~~ \verb"Z"$ ~~)                          \\
assignmentStmt   &$\lra$& varList = function $\verb"("$ varList  $\verb")"$                                                                     \\
varList &$\lra$& var \{$\verb","$ var  \}                                                                        \\
var  &$\lra$&   catVar ~~$|$~~ functorVar~~ $|$~~ ntVar~~\\
catVar &$\lra$& ${\bf 0}~~|~~ {\bf 1}~~|~~ \bf{2} ~~|~~\A ~~|~~ \B ~~|~~ \C \ldots$ \\ 
functorVar &$\lra$& $Id_{\bf 0} ~~|~~Id_{\bf 1} ~~|~~ Id_{\bf 2} ~~|~~ !_{\bf 01} ~~|~~ !_{\bf 02}  ~~|~~ !_{\bf 21} 
~~|~~F~~|~~G~~|~~H \cdots$\\ 
ntVar &$\lra $& $\alpha~~ |~~ \beta ~~|~~ \gamma \ldots$ \\   
function   &$\lra$&   
\verb"Source"$_1$~~$|$~~
\verb"Source"$_2$~~$|$~~
\verb"Target"$_1$~~$|$~~
\verb"Target"$_2$~~$|$~~\\&&
\verb"Ident"$_0$~~$|$~~
\verb"Ident"$_1$~~$|$~~
\verb"Op"$_0$~~$|$~~
\verb"OP"$_1$~~$|$~~ 
\verb"Comp"~~$|$~~
\verb"Hcomp"~~$|$~~
\verb"Vcomp"~~$|$~~\\&&
\verb"Pow"$_0$~~$|$~~
\verb"Pow"$_1$~~$|$~~
\verb"KanEx"~~$|$~~
\verb"KanExInd"~~$|$~~
\verb"KanLif"~~$|$~~
\verb"KanLifInd"~~$|$~~\\&&
\verb"MapComp" \\   
ifStmt  &$\lra$&   \verb"If " ntVar \verb" == " ntVar \verb" goto " label\\
return &$\lra$&   [label] \verb"Return(" varList\verb" )"
\end{tabular}

\vspace{.5in}

The basic types of Sammy are $\c$, $\f$, and $\n$ which correspond to categories, functors and natural transformations. There are maps between these types as follows:
$$\xymatrix{\n \ar@<.1in>[rrr]^s\ar@<-.1in>[rrr]_t&&& \f \ar@<.1in>[rrr]^s\ar@<-.1in>[rrr]_t&&& \c}$$  
These maps correspond to the source and targets of natural transformations and of functors.

In order to describe the operations in Sammy we need dependent type theory. Since the notation in the literature is not standardized we will steer clear of some of the notation that is harder to follow. For every two categories $\A$ and $\B$ there is a type $\f_{\A,\B}$ which corresponds to all the functors from $\A$ to $\B$. Similarly, for all pairs of functors $F:\A \lra \B$ and $G: \A \lra \B$ there is a type $\n_{\A; F,G;\B}$ that corresponds to all natural transformations from $F$ to $G$. Using the product types of these dependent types we can formally give the semantics of Sammy. 

The semantics of Sammy takes place in the universe $\Cat$, the 2-category of small categories and functors. 
The input, output and semantics of the Sammy operations are given in Table 1. For the semantics, we use the standard notation of \cite{CWM}. 

There are some other conventions about semantics that we follow (some are repeated here from the beginning of the paper for completion sake.)
\begin{enumerate}
\item  We have the following constant categories
\begin{itemize}
\item $\bf 0$, the empty category
\item $\bf 1$,  the category with one object, $0$,
\item  $\bf 2$, the category $0\lra 1$ with two objects and one nontrivial morphism between them
\item   \verb"Cat", the category of all small categories
\end{itemize}
\item  Kan extensions and Kan liftings are only defined up to a unique isomorphism. Hence, products, coproducts, equalizers, colimits, etc are also only defined up to a unique isomorphism. We leave it as implementation-dependent as to which of the many Kan extensions and Kan liftings Sammy returns.  Because of the universal property stated with the unique isomorphism, categorically they are all indistinguishable.
\item The program will move forward every line until there is a conditional statement. If the two natural transformations are equal, then the program will go to the first line in the program that has the given label. 
\end{enumerate} 

\begin{landscape}
\begin{table}
\caption{\bf Functions in the Sammy Programming Language}
\centering
\begin{tabular}{||l||c|c|l|l||}
\hline\hline
Function& Input & Output & Note&Semantics\\
\hline\hline
$\verb"Source"_1$&$\f$ & $\c$ & Source of the functor& $\A=\verb"Source"_1(F:\A\lra \B)$ \\
\hline
$\verb"Source"_2$&$\n$  &$\f$& Source of the N.T.& $F=\verb"Source"_2(\alpha: F\Lra G)$ \\
\hline
$\verb"Target"_1$&$\f$ & $\c$ & Target of the functor& $\B=\verb"Target"_1(F:\A\lra \B)$  \\
\hline
$\verb"Target"_2$&$\n$   &$\f$& Target of the N.T.& $G=\verb"Target"_2(\alpha: F\Lra G)$ \\
\hline
$\verb"Ident"_0$& $\c$  &$\f$& Identity functor  &$Id_{\A}=\verb"Ident"_0(\A)$\\
\hline
$\verb"Ident"_1$& $\f$ &$\n$ & Identity N.T.  &$\iota_{F}=\verb"Ident"_1(F:\A\lra \B)$ \\
\hline
$\verb"Op"_0$& $\c$  & $\c$ & Opposite category&$\A^{op}=\verb"Op"_0(\A)$ \\
\hline
$\verb"OP"_1$&$\f$  &$\f$& Opposite functor & $F^{op}=\verb"Op"_1(F:\A\lra \B) $ \\
\hline
$\verb"Comp"$& $\f_{\B,\C} \times \f_{\A,\B}$ &$\f_{\A,\C}$& Comp. of functors & $G\circ F=\verb"Comp"(G,F)$\\
\hline
$\verb"Hcomp"$&$\n_{\B;F',G';\C} \times  \n_{\A;F,G;\B}$  &$\n_{\A;F'\circ F,G' \circ G ;\C}$ &Horizontal comp. N.T.& $\alpha \circ \beta = \verb"Hcomp"(\alpha, \beta) $\\
\hline
$\verb"Vcomp"$&$\n_{\A;G,H;\B} \times \n_{\A;F,G;\B}$ & $\n_{\A;F,H;\B}$& Vertical comp. N.T.& $\sigma \bullet \tau=\verb"Vcomp"(\sigma, \tau) $\\
\hline
$\verb"Pow"_0$&  $\c \times \c$   & $\c$ & Power category & $\B^{\A}=\verb"Pow"_0(\A,\B) $ \\
\hline
$\verb"Pow"_1$&$\f_{\C,\D} \times \f_{\A,\B}$   &$\f_{\C^\B, \D^\A}$& Power functor& $G^F: \C^\B \lra \D^\A=\verb"Pow"_1(F,G)$ \\
\hline
$\verb"KanEx"$&$\f_{\C,\A}\times \f_{\C,\B}$ &$\f_{\A,\B}\times \n_{\C;Ran_F(G)\circ G,F;\A}$ & R. Kan extension & $(Ran_F(G),\alpha )=\verb"KanEx"(F,G) $\\
\hline
$\verb"KanExInd"$& $\f_{\C,\A}\times \f_{\C,\B}\times\f_{\A,\B}\times \n_{\A;H\circ G,F;\C}$ & $\n_{\A;H,Ran_F(G);\B}$& Induced by r. Kan ext.& $\gamma=\verb"KanExInd" (G,F,H,\beta)$\\
\hline
$\verb"KanLif"$&$\f_{\B,\C} \times \f_{\A,\C}$& $\f_{\A,\B}\times \n_{\A;P\circ RLif_P(F),F;\C}$ & R. Kan lifting & $(RLif_P(F), \alpha)=\verb"KanLif"(P,F) $\\
\hline
$\verb"KanLifInd"$& $\f_{\B,\C} \times \f_{\A,\C}\times \f_{\A,\B}\times \n_{\A;P\circ H,F;\C}$ & $\n_{\A;H,RLif_P(F);\B}$& Induced by r. Kan lifting& $\gamma=\verb"KanLifInd"(P,F, H, \beta) $\\
\hline
$\verb"MapComp"$ &  $\c$  & $\f$ & Composition functor& $(F:\C^{\st}\times_{\C^{\so}} \C^{\st} \rightarrow \C^{\st})=\verb"MapComp"(\C) $\\
\hline\hline
\end{tabular}
\end{table}
\end{landscape}

\end{document}